\documentclass[aps,twocolumn,superscriptaddress,pre]{revtex4}

\usepackage{graphicx}
\usepackage{amsmath}
\usepackage{amssymb}

\makeatletter
\def\oversortoftilde#1{\mathop{\vbox{\m@th\ialign{##\crcr\noalign{\kern3\p@}%
      \sortoftildefill\crcr\noalign{\kern3\p@\nointerlineskip}%
      $\hfil\displaystyle{#1}\hfil$\crcr}}}\limits}
\def\sortoftildefill{$\m@th \setbox\z@\hbox{$\braceld$}%
  \braceld\leaders\vrule \@height\ht\z@ \@depth\z@\hfill\braceru$}
\makeatother

\begin{document}
\title{Guiding-center motion for electrons in strong laser fields}

\author{J. Dubois}
\affiliation{Aix Marseille Univ, CNRS, Centrale Marseille, I2M, Marseille, France}
\author{S. A. Berman}
\affiliation{Aix Marseille Univ, CNRS, Centrale Marseille, I2M, Marseille, France}
\affiliation{School of Physics, Georgia Institute of Technology, Atlanta, Georgia 30332-0430, USA}
\author{C. Chandre}
\affiliation{Aix Marseille Univ, CNRS, Centrale Marseille, I2M, Marseille, France}
\author{T. Uzer}
\affiliation{School of Physics, Georgia Institute of Technology, Atlanta, Georgia 30332-0430, USA}

\begin{abstract}
We consider the dynamics of electrons in combined strong laser and Coulomb fields. Under a time-scale separation condition, we reduce this dynamics to a guiding-center framework. More precisely, we derive a hierarchy of models for the guiding-center dynamics based on averaging over the fast motion of the electron using Lie transforms. 
The reduced models we obtain describe well the different ionization channels, in particular, the conditions under which an electron is rescattered by the ionic core or is directly ionized. The comparison between these models highlights the models which are best suited for a qualitative and quantitative agreement with the parent dynamics. 
\end{abstract}


\maketitle

\section{Introduction}

Electrons in atoms or molecules subjected to laser fields can tunnel ionize when the laser intensity is strong enough to compete with the Coulomb attraction~\citep{Augst1989,Corkum2007,Agostini2008,Krausz2009}. After tunneling, the electron can for example ionize directly, or be rescattered by the ionic core~\citep{Corkum1993,Schafer1993,Krausz2009}. Rescattering occurs when the electron tunnel ionizes, then comes back to the ionic core region, and is finally ionized after a strong interaction with the ionic core. This process, also referred to as a recollision, is the keystone of strong field physics~\citep{Becker2008}. In contrast, direct ionization occurs when the electron never comes back to the ionic core after tunnel ionization. The parameters of the laser pulse, e.g., its intensity, frequency, polarization, and envelope, play a paramount role in the ionization process, as reflected by the shaping in the momentum distributions of the ionized electrons measured at the detectors~\citep{Becker2002,Li2013,Landsman2013,Li2017}. In order to better analyze and interpret what is measured at the detectors, there is a need to better understand the motion of the electrons after tunneling~\citep{Zuo1996,Meckel2008,Peters2011,Huismans2011,Blaga2012,Kerbstadt2017}. 
\par
To describe the motion of the electron after tunneling, a convenient approximation is usually performed: Since the electron is relatively far from the ionic core after tunneling and the electric field is strong, the Coulomb interaction is neglected. This leads to the so-called strong field approximation~\citep{Corkum1993} (SFA), which is one of the main theoretical tools in strong-field atomic physics. The main advantage of this approximation is that it allows the explicit computation of the trajectories of the electron since the equations of motion are linear. However, the SFA often leads to disagreements or misleading interpretations when it is confronted with experimental data~\citep{Brabec1996,Goreslavski2004,Landsman2013,Gillen2001}. For example, in linearly polarized (LP) fields, the SFA suggests that if the electron does not return to the ionic core within one laser cycle after tunneling, it never comes back to the ionic core at all. However, multiple laser cycle recollisions are essential for the quantitative agreement between theories and experiments in non-sequential double ionization (NSDI)~\citep{Brabec1996, Bhardwaj2001, Yudin2001_1, Yudin2001_2, Yudin2001_3}. In circularly polarized (CP) fields, the SFA suggests that the drift-velocity of the electron pushes it away from the ionic core, without rescattering. The predicted absence of recollisions is in contradiction with the knee structure observed in the double ionization probability curves, a signature of the recollision process, observed in simulations~\citep{Mauger2010_PRL,Chen2017,Fu2012} and experiments~\citep{Gillen2001}. In fact, multiple laser cycle recollisions or recollisions in CP become possible only when the Coulomb interaction is taken into account. In addition, Coulomb effects such as Coulomb focusing~\citep{Brabec1996,Comtois2005} and Coulomb asymmetry~\citep{Goreslavski2004,Bandrauk2000} are measurable and significant in above-threshold ionization (ATI)~\citep{Comtois2005, Landsman2013, Dubois2018}. Therefore, the Coulomb potential cannot be ignored, even far away from the ionic core.
\par
In the length gauge (see Ref.~\citep{Peng2015} for a review), the Hamiltonian of an electron interacting with its parent ion and a laser field reads 
\begin{equation}
\label{eq:main_hamiltonian}
H ( \mathbf{r}, \mathbf{p},t )= \dfrac{|\mathbf{p}|^2}{2} + V(\mathbf{r}) + \mathbf{r} \cdot \mathbf{E}(t) ,
\end{equation} 
where $\mathbf{r}$ and $\mathbf{p}$ are the position of the electron and its canonically conjugate momentum, respectively. Atomic units (a.u.) are used unless stated otherwise. Here, we have considered a single-active electron atom, where the effective charge is equal to one due to the shielding effect by the inner electrons. Furthermore, the timescale of interest is of the order of femtoseconds. During this short timescale, the ionic core is too massive to display relevant variations of its position, and consequently, we have considered the ionic core to be static. Moreover, the characteristic distance of the electron in the laser field is small compared to the wavelength of the laser field, and consequently, we have neglected the spatial dependence of the electric field (dipole approximation). The polarization of the electric field is crucial in the analysis of the variety of nonlinear phenomena~\citep{Antoine1996,Goreslavski1996,Shafir2013,Landsman2013_NJP,Hofmann2014,Dimitrovski2015,He2015,Chen2017,Danek2018,Maurer2018}. The elliptically polarized electric field is 
\begin{equation}
\label{eq:electric_field}
\mathbf{E}(t) = f(t) \dfrac{E_0}{\sqrt{\xi^2 + 1}} \left[ \hat{\mathbf{x}} \cos ( \omega t + \phi ) + \hat{\mathbf{y}} \xi \sin ( \omega t + \phi ) \right] .
\end{equation} 
The laser ellipticity is $\xi$, where in LP and CP fields, $\xi = 0$ and $\xi = 1$, respectively. The laser amplitude in a.u. is given by $E_0 \approx 5.338 \times 10^{-9} \sqrt{I}$, with $I$ the laser intensity in $\mathrm{W}\cdot \mathrm{cm}^{-2}$. The laser phase at $t=0$ is $\phi$. In this manuscript, we consider the laser envelope to be $f(t) = 1$. Here, we consider a single frequency $\omega$ for the laser field. 

\begin{figure}
\centering
\includegraphics[width=0.5\textwidth]{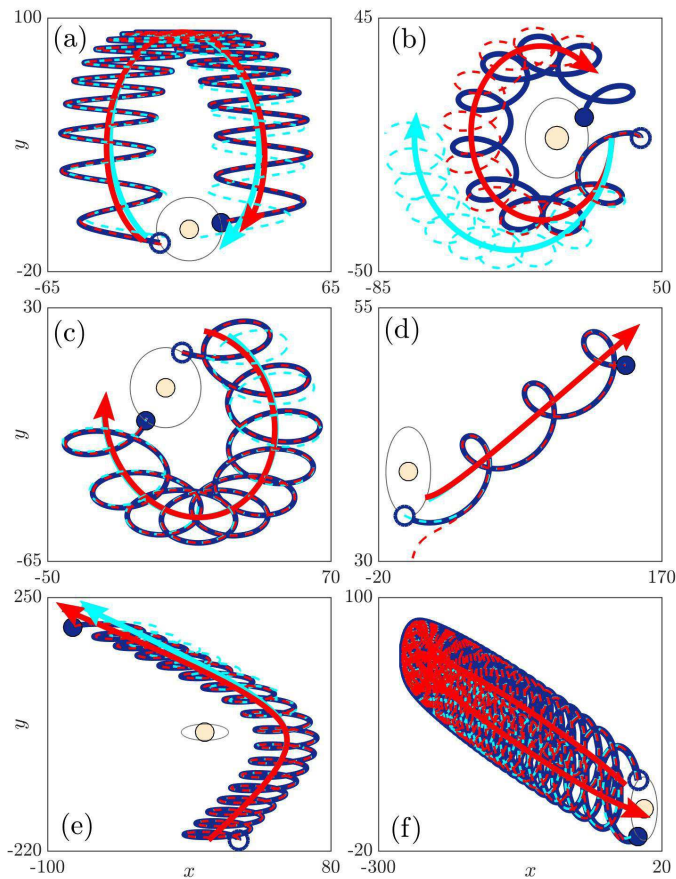}
\caption{Typical electron trajectories for $I = 10^{14} \; \mathrm{W}\cdot\mathrm{cm}^{-2}$, $d=2$ and $\omega = 0.05 \; \mathrm{a.u.}$ in the polarization plane $(x,y)$. The ellipticities are: (a) $\xi=0$, (b--e) $\xi = 0.5$, (f) $\xi = 1$. The dark blue curve is the electron trajectory of Hamiltonian~\eqref{eq:main_hamiltonian}. The light blue and red curves are the trajectories of the models $\mathrm{G}_2$ and $\mathrm{G}_5$, respectively, with initial conditions far from the ionic core. For each model, the solid and dashed curves are the guiding-center and the approximate trajectories, respectively. The lightly shaded circle is the position of the ionic core at the origin, and the black circle surrounding the origin is $|\mathbf{r}| = 15 \; \mathrm{a.u.}$ All quantities are in atomic units.}
\label{fig:trajectories}
\end{figure}

Figure~\ref{fig:trajectories} displays six typical trajectories of Hamiltonian~\eqref{eq:main_hamiltonian}. We notice that these trajectories display fast oscillations around a mean trajectory, which we call the guiding-center trajectory. In this article, we use this clear separation of scales to derive models for the guiding-center dynamics. 
\par
Specifically, we average Hamiltonian~\eqref{eq:main_hamiltonian} over the fast motion using Lie transforms in order to simplify the electron dynamics and clearly distinguish the different ionization channels for the electron. As a consequence, we derive a hierarchy of averaged models which fully take into account the Coulomb potential. In these models, the electron in the combined strong laser and Coulomb fields follows a guiding center trajectory~\citep{Dubois2018}. Actually, there are several possible guiding centers, depending on the order up to which the averaging is performed. Computations are presented for a $d$-dimensional configuration space, where $d=1$ (only for linear polarization) or $d=2,3$. The objective of this manuscript is to derive and investigate these reduced models. 
\par
In Sec.~\ref{sec:derivation_of_the_hierarchy}, we describe the procedure we use for averaging Hamiltonian~\eqref{eq:main_hamiltonian} over the fast time scale, and we derive a hierarchy of models for the guiding center dynamics. In Sec.~\ref{sec:hierarchy_models}, we first compare these reduced models with the dynamics associated with Hamiltonian~\eqref{eq:main_hamiltonian}. In particular, we show the relevance of two models in the hierarchy. Then we analyze the dynamics of the guiding-center models in phase space, highlighting regular and chaotic regions and their relation with the trajectories of Hamiltonian~\eqref{eq:main_hamiltonian}.

\section{Derivation of the hierarchy of models \label{sec:derivation_of_the_hierarchy}}
\subsection{Reminders on canonical Lie transforms}
In this section, we recall some basic features on canonical transformations in the framework of canonical Lie transforms. For more details, we refer to Refs.~\cite{Cary1981,Cary1983}.
We consider a Hamiltonian system with phase-space variables $\mathbf{z}$, a Hamiltonian $H({\bf z})$ and a Poisson bracket $\lbrace \cdot , \cdot \rbrace$ (whether it is a canonical or non-canonical bracket). Canonical Lie transforms are near-identity changes of coordinates $\mathbf{z} \mapsto \bar{\mathbf{z}} (\mathbf{z})$, generated by a scalar function $S(\mathbf{z})$, called the generating function, and given by
\begin{eqnarray}
\bar{\bf z}&=&\exp \left( -\mathcal{L}_S \right) {\bf z},\label{eq:Change_coordinates_Lie_transform}\\
&=& {\bf z}-\lbrace S , {\bf z} \rbrace + \dfrac{1}{2} \lbrace S , \lbrace S , {\bf z} \rbrace \rbrace  + \ldots, \nonumber
\end{eqnarray}
where $\mathcal{L}_S$ is the Liouville operator defined by $\mathcal{L}_S F = \lbrace S , F \rbrace$. Canonical Lie transforms have several properties:
\begin{enumerate}
\item[(i)]
$F(\exp (\mathcal{L}_S) \mathbf{z}) = \exp (\mathcal{L}_S) F (\mathbf{z})$, which comes from the Leibniz rule,
\item[(ii)]
$\lbrace \exp (\mathcal{L}_S) F, \exp (\mathcal{L}_S) G \rbrace = \exp (\mathcal{L}_S) \lbrace F, G\rbrace$, which comes from the Jacobi identity and the antisymmetry of the Poisson bracket,
\end{enumerate}  
for any scalar functions $F(\mathbf{z})$ and $G(\mathbf{z})$. As a consequence of Property (i) and of the scalar invariance $\bar{F}(\bar{\bf z})=F({\bf z})$, these changes of variables
modify any observable $F({\bf z})$, and in particular the Hamiltonian $H({\bf z})$, into
\begin{eqnarray}
\label{eq:Observable_transformation_Lie_transform}
\bar{F} (\bar{\mathbf{z}}) &=& \exp \left( \mathcal{L}_S \right) F (\bar{\mathbf{z}}).
\end{eqnarray}
Property (ii) ensures that these changes of coordinates do not affect the expression of the Poisson bracket, i.e., they are canonical transformations. One significant advantage of these transformations is that they are explicit functions and they can be easily inverted: ${\mathbf{z}} = \exp (\mathcal{L}_S) \bar{\mathbf{z}}$. This way, we can recover all the information on the particle dynamics from the transformed (averaged) system.
\par
These canonical Lie transforms are particularly well suited for perturbation theory. 
If the Hamiltonian is of the form $H = H_0 + \epsilon W$, where $H_0$ is the Hamiltonian of the unperturbed system, $W$ is the perturbation and $\epsilon$ is an ordering (small) parameter, a canonical Lie transform generated by a generating function $S(\bar{\bf z})$ (of order $\epsilon$), applied to $H$ is able to remove the unwanted part of the perturbation, called $\widetilde{W}$, and move its influence to higher orders in $\epsilon$. More explicitly, at the lowest order, the expression of the new Hamiltonian expressed in the new variables $\bar{\bf z}$ is
\begin{eqnarray}
\label{eq:Lie_transform_Hamiltonian_special_form}
\bar{H} &=& \exp \left( \mathcal{L}_S \right) H \\
			&=& H_0 + \epsilon  W + \lbrace S, H_0 \rbrace \nonumber \\
			&&+ \epsilon \lbrace S, W \rbrace + \dfrac{1}{2}  \lbrace S, \lbrace S, H_0 \rbrace\rbrace  +  \ldots. \nonumber
\end{eqnarray} 
Choosing appropriately the generating function $S$ such that $\lbrace S, H_0 \rbrace = - \epsilon \widetilde{W}$, unwanted terms in the perturbation $W$ can be pushed from order $\epsilon$ to order $\epsilon^2$, meaning that the order $\epsilon$ in the Hamiltonian becomes $W-\widetilde{W}$. For instance, one can suppress fast oscillating terms contained in $W$. Then, the associated canonical change of coordinates is determined using Eq.~\eqref{eq:Change_coordinates_Lie_transform}.

\subsection{Averaging the electron dynamics}
We notice that these transformations are defined for autonomous systems. Hamiltonian~\eqref{eq:main_hamiltonian} has an explicit time dependence through the electric field. 
Therefore, we first increase phase space to include time $t$, and consider its canonically conjugate variable $k$. The extended Hamiltonian~\eqref{eq:main_hamiltonian} becomes
\begin{equation}
\label{eq:main_Hamiltonian_autonomous}
H (\mathbf{r}, \mathbf{p}, t, k) = k + \dfrac{|\mathbf{p}|^2}{2} + V(\mathbf{r})  + \mathbf{r} \cdot \mathbf{E} (t) .
\end{equation}
The extended Poisson bracket is
\begin{equation}
\lbrace F, G \rbrace = \dfrac{\partial F}{\partial \mathbf{r}} \cdot \dfrac{\partial G}{\partial \mathbf{p}} - \dfrac{\partial F}{\partial \mathbf{p}} \cdot \dfrac{\partial G}{\partial \mathbf{r}} + \dfrac{\partial F}{\partial t} \dfrac{\partial G}{\partial k} - \dfrac{\partial F}{\partial k} \dfrac{\partial G}{\partial t} ,
\end{equation}
where the operators $\partial/\partial\mathbf{r} = (\partial/\partial x, \partial/\partial y, \partial /\partial z)$ and $\partial/\partial\mathbf{p} = (\partial /\partial p_x, \partial/\partial p_y, \partial / \partial p_z)$. 
\par
The hypothesis we make for the derivation of our hierarchy of reduced models is that the characteristic time of the ionized electron trajectory is large compared to a laser cycle $T = 2\pi/\omega$, i.e., $\omega \mapsto \omega/\epsilon$ where $\epsilon$ is an ordering parameter which is explicitly introduced for bookkeeping purposes. Performing the canonical change of coordinates $\bar{t} = t/\epsilon$ and $\bar{k} = \epsilon k$, and re-scaling the energy, Hamiltonian~\eqref{eq:main_Hamiltonian_autonomous} becomes
\begin{equation}
\label{eq:Hamiltonian_H0}
H^{(0)} (\mathbf{r}, \mathbf{p}, t , k; \epsilon) = k + \epsilon \left[ \dfrac{|\mathbf{p}|^2}{2} + V(\mathbf{r}) + \mathbf{r}\cdot \mathbf{E}(t) \right]  ,
\end{equation}
where we have removed the bars in the new variables. We apply canonical Lie transforms as described above in order to perform the averaging of Hamiltonian~\eqref{eq:Hamiltonian_H0} over the fast time scale, by pushing time-dependent terms in the Hamiltonian to higher order terms in $\epsilon$. 

\subsubsection{Gauge velocity transformation \label{app:Gauge_velocity_transformation}}
As an example, we consider the transformation from the length gauge to the velocity gauge~\citep{Peng2015} in Hamiltonian~\eqref{eq:Hamiltonian_H0}. It is given by the following change of coordinates
\begin{eqnarray*}
\bar{\mathbf{r}} &=& \mathbf{r} , \\
\bar{\mathbf{p}} &=& \mathbf{p} - \epsilon \mathbf{A} (t) ,
\end{eqnarray*}
where $\mathbf{A}(t)$ is the vector potential defined by $\mathbf{E}(t) = - \partial \mathbf{A}(t) / \partial t$. This transformation is a canonical change of coordinates which can be formulated as a canonical Lie transform generated by
\begin{equation}
\label{eq:app_S1}
S^{(1)} = \epsilon \, \mathbf{r} \cdot \mathbf{A}(t).
\end{equation}
The Hamiltonian in the velocity-gauge coordinates becomes
\begin{eqnarray}
\label{eq:Hamiltonian_H1}
{H}^{(1)} &=& \exp (\mathcal{L}_{S^{(1)}}) H^{(0)}\\
&=& \bar{k} + \epsilon \left[ \dfrac{1}{2} \left(\bar{\mathbf{p}} + \epsilon \mathbf{A}(t) \right)^2 +  V(\bar{\mathbf{r}}) \right]  , \nonumber \\
		&=&   \bar{k} + \epsilon \left[ \dfrac{|\bar{\mathbf{p}}|^2}{2} + V(\bar{\mathbf{r}}) \right] + \epsilon^2 \bar{\mathbf{p}} \cdot \mathbf{A}(t) \nonumber + \epsilon^3 \dfrac{\mathbf{A}^2(t)}{2} .
\end{eqnarray}
We observe that the time-dependence in Hamiltonian~\eqref{eq:Hamiltonian_H0}, present in ${\bf E}(t)$, is of order $\epsilon$, while in Hamiltonian~\eqref{eq:Hamiltonian_H1} this time-dependence is moved to order $\epsilon^2$. 

\subsubsection{Iterative procedure \label{sec:iterative_procedure}}
We iterate the above-procedure to higher order in $\epsilon$. 
We assume that after the $N$-th step of the procedure, all the time-dependent terms in the present averaged Hamiltonian are removed up to order $\epsilon^N$, that is, the time-dependence of the averaged Hamiltonian is of order $\epsilon^{N+1}$. We assume that the total generating function up to order $\epsilon^{N}$ is known. The total generating function at this step is
\begin{equation}
\label{eq:S_N_sum_definition}
S^{(N)}({\bf r},{\bf p},t;\epsilon) = \sum_{n = 1}^{N} \epsilon^n S_n ({\bf r},{\bf p},t) , 
\end{equation}
and the corresponding averaged Hamiltonian is denoted
\begin{eqnarray*}
{H}^{(N)} &=& \exp (\mathcal{L}_{S^{(N)}}) H^{(0)}\\
 &=& k + \sum_{n=1}^{N} \epsilon^n h_n({\bf r},{\bf p})+\epsilon^{N+1} R_{N+1}({\bf r},{\bf p},t;\epsilon),
\end{eqnarray*}
where $h_n(\mathbf{r}, \mathbf{p})$ are the coefficients in the series expansion of the Hamiltonian that no longer depends on time, while $R_{N+1}({\bf r},{\bf p},t;\epsilon)$ is the remainder of the Hamiltonian which still depends on time. The objective of the iterative method is to find the modified generating function $S^{(N+1)}$ [which amounts to finding the extra function $S_{N+1}$ in Eq.~\eqref{eq:S_N_sum_definition}] to remove the time-dependence in the term $R_{N+1}$ at the lowest order. 
\par  
The averaged Hamiltonian ${H}^{(N+1)}$ whose time-dependence is of order $\epsilon^{N+2}$ is
\begin{eqnarray*}
{H}^{(N+1)} &=& \exp (\mathcal{L}_{S^{(N+1)}}) H^{(0)} , \nonumber \\
						&=& \exp (\epsilon^{N+1} \mathcal{L}_{S_{N+1}}) \exp \left( \mathcal{L}_{S^{(N)}} \right) {H}^{(0)} + O(\epsilon^{N+2}) , \nonumber \\
						&=& \exp (\epsilon^{N+1} \mathcal{L}_{S_{N+1}}) {H}^{(N)} + O(\epsilon^{N+2}) , \nonumber \\
						&=& H^{(N)} + \epsilon^{N+1} \left( R_{N+1} + \dfrac{\partial S_{N+1}}{\partial t} \right) + O(\epsilon^{N+2}) .
\end{eqnarray*}
The time-fluctuating terms in $R_{N+1}$ are denoted $\oversortoftilde{R_{N+1}}$, and are defined by
\begin{equation*}
\oversortoftilde{R_{N+1}}=R_{N+1}-\frac{1}{T}\int_0^{T} {\rm d}t R_{N+1}.
\end{equation*}
In order to eliminate the time-fluctuating terms at order $\epsilon^{N+1}$, the component $S_{N+1}$ of the generating function $S^{(N+1)}$ is chosen as
\begin{equation}
\label{eq:S_iteratve_procedure}
S_{N+1} = - \int {\rm d}t \oversortoftilde{R_{N+1}({\bf r},{\bf p},t;0)} ,
\end{equation}
where the primitive is chosen such that the mean value of $S_{N+1}$ with respect to $t \in [0,T]$ is zero. At each step the functions $R_{n}$ have to be computed up to order $M$ where $M$ is the last order for which the averaged Hamiltonian will be computed analytically. We perform these computations using a symbolic computation software.  

\subsubsection{Averaged Hamiltonians \label{app:summary_transformations}}
We apply the above-described procedure to push the time-dependence in Hamiltonian~\eqref{eq:Hamiltonian_H0} to order $\epsilon^8$ using ${H}^{(7)} = \exp (\mathcal{L}_{S^{(7)}} ) H^{(0)}$. Below we provide the explicit expression for $S^{(6)}$. The higher-order components are too lengthy to report and their expressions are not particularly enlightening. 
\begin{widetext}
\begin{eqnarray}
\label{eq:app_total_generator_function}
S^{(6)} &=& \epsilon \, \mathbf{r} \cdot \mathbf{A}(t) - \dfrac{\epsilon^2}{\omega^2} \, \mathbf{p} \cdot \mathbf{E}(t) + \dfrac{\epsilon^3}{\omega^2} \mathbf{A}(t) \cdot  \left( \dfrac{\mathbf{E}(t)}{4}  + \dfrac{\partial}{\partial \mathbf{r}} \right) V - \dfrac{\epsilon^4}{\omega^4} \mathbf{p} \cdot \dfrac{\partial}{\partial \mathbf{r}} \left( \mathbf{E}(t) \cdot \dfrac{\partial}{\partial \mathbf{r}} \right) V \nonumber \\
	&& + \dfrac{\epsilon^5}{\omega^4} \left[ \left( \dfrac{\mathbf{E}(t)}{4} +  \dfrac{\partial V}{\partial \mathbf{r}} \right) \cdot \dfrac{\partial}{\partial \mathbf{r}} - \left( \mathbf{p} \cdot \dfrac{\partial}{\partial \mathbf{r}} \right)^2 \right] \left( \mathbf{A}(t) \cdot \dfrac{\partial}{\partial \mathbf{r}} \right) V  - \dfrac{5 \epsilon^6}{8 \omega^6} \mathbf{p}\cdot \dfrac{\partial}{\partial \mathbf{r}} \oversortoftilde{\left( \mathbf{E}(t) \cdot \dfrac{\partial}{\partial \mathbf{r}} \right)^{2}} V \nonumber \\
			&& - \dfrac{\epsilon^6}{\omega^6} \left[ \left( \mathbf{p} \cdot \dfrac{\partial}{\partial \mathbf{r}} \right) \left( \dfrac{\partial V}{\partial \mathbf{r}} \cdot \dfrac{\partial}{\partial \mathbf{r}} \right) - \left(\mathbf{p} \cdot \dfrac{\partial}{\partial \mathbf{r}} \right)^3 + 2\left( \dfrac{\partial V}{\partial \mathbf{r}} \cdot \dfrac{\partial}{\partial \mathbf{r}} \right) \left(\mathbf{p} \cdot \dfrac{\partial}{\partial \mathbf{r}} \right) \right] \left( \mathbf{E}(t) \cdot \dfrac{\partial}{\partial \mathbf{r}} \right) V .
\end{eqnarray}
\end{widetext}
Here, we have used the fact that the electric field is monochromatic and satisfies $\omega^2 \mathbf{E}(t) = - \partial^2 \mathbf{E} / \partial t^2$. The averaged Hamiltonian ${H}^{(7)}$ is
\begin{widetext}
\begin{eqnarray}
\label{eq:total_averaged_Hamiltonian}
{H}^{(7)} &=& \bar{k} + \epsilon \left[ \dfrac{|\bar{\mathbf{p}}|^2}{2} + V ( \bar{\mathbf{r}} )   \right] + \epsilon^3 \mathrm{U}_p + \epsilon^5 \dfrac{\mathrm{U}_p}{\omega^2 (\xi^2+1)} \left( \dfrac{\partial^2 V}{\partial x^2} + \xi^2 \dfrac{\partial^2 V}{\partial y^2}\right) \nonumber \\ 
			&&+ \epsilon^7 \dfrac{\mathrm{U}_p}{\omega^4(\xi^2+1)} \left[ \left| \dfrac{\partial}{\partial \mathbf{r}} \left( \dfrac{\partial V}{\partial x} \right) \right|^2 + \xi^2 \left| \dfrac{\partial}{\partial \mathbf{r}} \left( \dfrac{\partial V}{\partial y} \right) \right|^2 \right] + O(\epsilon^8)  ,
\end{eqnarray}
\end{widetext}
where all the derivatives are evaluated at $\bar{\mathbf{r}}$. By truncating the Hamiltonian at a given order, we notice that the reduced (time-independent) Hamiltonians up to order $\epsilon^7$ are of the form 
\begin{equation*}
{H}(\bar{\bf r},\bar{\bf p})= \frac{\vert \bar{\bf p}\vert^2}{2}+V_{\rm eff}(\bar{\bf r}) ,
\end{equation*}
with an effective potential $V_{\rm eff}$. In particular, this highlights a particular property in the reduction process that the reduction procedure does not generate $\bar{\mathbf{p}}$-dependent terms in the Hamiltonian other than the kinetic energy, up to order $\epsilon^7$. At order $\epsilon^8$, the term which is generated in $H^{(8)}$ is linear in the momenta; therefore, it can easily be eliminated by a translation in $\bar{\bf p}$ (which is a canonical transformation). At order $\epsilon^9$, the terms which are generated are quadratic in the momenta $\bar{\bf p}$, and we do not consider these terms in what follows. Therefore, our analysis is valid up to order $\epsilon^8$. In such a way, it is particularly convenient to define effective potentials. Depending on the order of truncation, we obtain three reduced Hamiltonians
\begin{widetext}
\begin{subequations}
\label{eq:Hamiltonians_H2_H5_H7}
\begin{eqnarray}
{H}_2 (\bar{\mathbf{r}}, \bar{\mathbf{p}}) &=& \frac{\vert \bar{\bf p}\vert^2}{2}+V(\bar{\bf r}), \label{eq:Hamiltonian_H2} \\
{H}_5 (\bar{\mathbf{r}}, \bar{\mathbf{p}}) &=& \frac{\vert \bar{\bf p}\vert^2}{2}+V(\bar{\bf r}) + \dfrac{\mathrm{U}_p}{\omega^2 (\xi^2+1)} \left( \dfrac{\partial^2 V}{\partial x^2} + \xi^2 \dfrac{\partial^2 V}{\partial y^2}\right), \label{eq:Hamiltonian_H5} \\
{H}_7 (\bar{\mathbf{r}}, \bar{\mathbf{p}}) &=& \frac{\vert \bar{\bf p}\vert^2}{2}+ V(\bar{\bf r}) + \dfrac{\mathrm{U}_p}{\omega^2 (\xi^2+1)} \left( \dfrac{\partial^2 V}{\partial x^2} + \xi^2 \dfrac{\partial^2 V}{\partial y^2}\right) + \dfrac{\mathrm{U}_p}{\omega^4 (\xi^2+1)} \left[ \left| \dfrac{\partial}{\partial \mathbf{r}} \left( \dfrac{\partial V}{\partial x} \right) \right|^2 + \xi^2 \left| \dfrac{\partial}{\partial \mathbf{r}} \left( \dfrac{\partial V}{\partial y} \right) \right|^2 \right] , \label{eq:Hamiltonian_H7}
\end{eqnarray}
\end{subequations}
\end{widetext}
where we have removed the small parameter $\epsilon$ which was originally introduced for bookkeeping purposes. Each of these Hamiltonians describes the dynamics of the guiding center at a different level of approximation. As a result of averaging, the Hamiltonians~\eqref{eq:Hamiltonians_H2_H5_H7} are conserved, in contrast to Hamiltonian~\eqref{eq:main_hamiltonian}. We notice that the quantities of the effective potentials depend on the main parameters of the electric field, its amplitude and its ellipticity, with the exception of $H_2$.
\par
The corresponding change of coordinates which maps $H^{(0)} ({\bf r},{\bf p},t,k;\epsilon)$ to ${H}^{(7)}$ is given by Eq.~\eqref{eq:Change_coordinates_Lie_transform} and its series expansion up to order $\epsilon^6$ is given by
\begin{widetext}
\begin{subequations}
\label{eq:total_coordinate_transformation}
\begin{eqnarray}
\bar{\mathbf{r}} &=&  \mathbf{r} - \dfrac{\epsilon^2}{\omega^2} \mathbf{E}(t) - \dfrac{\epsilon^4}{\omega^4} \dfrac{\partial}{\partial \mathbf{r}} \left( \mathbf{E}(t) \cdot \dfrac{\partial}{\partial \mathbf{r}} \right) V - \dfrac{2\epsilon^5}{\omega^4} \dfrac{\partial}{\partial \mathbf{r}} \left( \mathbf{p} \cdot \dfrac{\partial}{\partial \mathbf{r}} \right) \left( \mathbf{A}(t) \cdot \dfrac{\partial}{\partial \mathbf{r}} \right) V - \dfrac{\epsilon^6}{8 \omega^6} \dfrac{\partial}{\partial \mathbf{r}} \oversortoftilde{ \left( \mathbf{E}(t) \cdot \dfrac{\partial}{\partial \mathbf{r}} \right)^2} V \nonumber \\ 
				&& - \dfrac{\epsilon^6}{\omega^6} \dfrac{\partial}{\partial \mathbf{r}} \left[ \dfrac{\partial V}{\partial \mathbf{r}} \cdot \dfrac{\partial}{\partial \mathbf{r}} - 3 \left( \mathbf{p} \cdot \dfrac{\partial}{\partial \mathbf{r}} \right)^2 \right] \left( \mathbf{E}(t) \cdot \dfrac{\partial}{\partial \mathbf{r}} \right) V - \dfrac{2 \epsilon^6}{\omega^6} \left( \dfrac{\partial V}{\partial \mathbf{r}} \cdot \dfrac{\partial}{\partial \mathbf{r}} \right) \left( \mathbf{E}(t) \cdot \dfrac{\partial}{\partial \mathbf{r}} \right) \dfrac{\partial V}{\partial \mathbf{r}}  + O(\epsilon^7) , \\
\bar{\mathbf{p}} &=& \mathbf{p} - \epsilon \mathbf{A}(t) - \dfrac{\epsilon^3}{\omega^2} \dfrac{\partial}{\partial \mathbf{r}} \left( \mathbf{A}(t) \cdot \dfrac{\partial}{\partial \mathbf{r}} \right) V + \dfrac{\epsilon^4}{\omega^4} \dfrac{\partial}{\partial \mathbf{r}} \left( \mathbf{p} \cdot \dfrac{\partial}{\partial \mathbf{r}} \right) \left( \mathbf{E}(t) \cdot \dfrac{\partial}{\partial \mathbf{r}} \right) V  \nonumber \\
		&-& \dfrac{\epsilon^5}{\omega^4} \dfrac{\partial}{\partial \mathbf{r}} \left[ \left( \dfrac{\mathbf{E}(t)}{4} + \dfrac{\partial V}{\partial \mathbf{r}} \right) \cdot \dfrac{\partial}{\partial \mathbf{r}} - \left( \mathbf{p} \cdot \dfrac{\partial}{\partial \mathbf{r}} \right)^2 \right] \left( \mathbf{A}(t) \cdot \dfrac{\partial}{\partial \mathbf{r}} \right) V + \dfrac{\epsilon^6}{8 \omega^6} \left( \mathbf{p}\cdot \dfrac{\partial}{\partial \mathbf{r}} \right) \oversortoftilde{\left( \mathbf{E}(t) \cdot \dfrac{\partial}{\partial \mathbf{r}} \right)^{2}} \dfrac{\partial V}{\partial \mathbf{r}} \nonumber \\
			&& + \dfrac{\epsilon^6}{\omega^6} \dfrac{\partial}{\partial \mathbf{r}} \left[ \left( \mathbf{p} \cdot \dfrac{\partial}{\partial \mathbf{r}} \right) \left( \dfrac{\partial V}{\partial \mathbf{r}} \cdot \dfrac{\partial}{\partial \mathbf{r}} \right) - \left(\mathbf{p} \cdot \dfrac{\partial}{\partial \mathbf{r}} \right)^3 + 2\left( \dfrac{\partial V}{\partial \mathbf{r}} \cdot \dfrac{\partial}{\partial \mathbf{r}} \right) \left(\mathbf{p} \cdot \dfrac{\partial}{\partial \mathbf{r}} \right) \right] \left( \mathbf{E}(t) \cdot \dfrac{\partial}{\partial \mathbf{r}} \right) V + O(\epsilon^7) .
\end{eqnarray}
\end{subequations}
\end{widetext}
If one needs to know the averaged Hamiltonian and/or the system of coordinates at the order $\epsilon^N$, one needs to truncate the $O(\epsilon^{N+1})$ terms of Eqs.~\eqref{eq:total_averaged_Hamiltonian} and/or~\eqref{eq:total_coordinate_transformation}, respectively. The expressions~\eqref{eq:total_coordinate_transformation} are used below to reconstruct the trajectories of the electrons from the trajectories of the guiding centers.  
\par
In summary, the hierarchy of models is composed of a time-independent Hamiltonian $H_m$ [see Eqs.~\eqref{eq:Hamiltonians_H2_H5_H7}] and a transformation from the electron coordinates to the guiding centers: 
\begin{equation}
\label{eq:transformation_Phi_n}
\Phi_n : (\mathbf{r}, \mathbf{p}) \mapsto (\bar{\mathbf{r}}, \bar{\mathbf{p}}) , 
\end{equation}
whose truncated expressions are given by the truncations of Eqs.~\eqref{eq:total_coordinate_transformation}.
The orders $m$ and $n$ refer to the order of the Hamiltonian and the transformation, respectively, after truncation of the perturbative expansion. We refer to the model 
\begin{equation*}
\mathrm{G}_n = ({H}_m,\Phi_n) ,
\end{equation*}
as the $n$-th order guiding-center model, where $n \geq m$, and we show the set of $(m,n)$ in Table~\ref{tab:set_couple_mn}. 

\begin{table}
\resizebox{0.7\columnwidth}{!}{
\begin{tabular}{ c @{\hspace{0.5cm}} c  c  c  c  c  c }
\hline\hline
			& $\Phi_2$ 	& $\Phi_3$	& $\Phi_4$	& $\Phi_5$	& $\Phi_6$	& $\Phi_7$ 	\\
\hline
 ${H}_2$		& $\mathrm{G}_2$ 	& $\mathrm{G}_3$ & $\mathrm{G}_4$ &  $\circ$   & $\circ$	&$\circ$  \\
 ${H}_5$    & $\circ$  	& $\circ$	& $\circ$	& $\mathrm{G}_5$ 	& $\mathrm{G}_6$ & $\circ$  \\
 ${H}_7$    & $\circ$	& $\circ$	& $\circ$	&  $\circ$	&  $\circ$  & $\mathrm{G}_7$  \\
\hline\hline
\end{tabular}}
\caption{The guiding-center models $\mathrm{G}_n = (H_m, \Phi_n)$, where $n \geq m$.}
\label{tab:set_couple_mn}
\end{table}


\subsubsection{Links with the Kramers-Henneberger potential}
At order $\epsilon^5$, our model is linked to the Kramers-Henneberger (KH) treatment of the motion of a charged particle in an external time-periodic electric field~\citep{Corso1995,Smirnova2000,Morales2011}. In a nutshell, the classical KH theory amounts to performing a canonical Lie transform generated by 
\begin{equation*}
S^{(2)} = \epsilon \, \mathbf{r} \cdot \mathbf{A}(t) - \dfrac{\epsilon^2}{\omega^2} \, \mathbf{p} \cdot \mathbf{E}(t),
\end{equation*}
on Hamiltonian~\eqref{eq:Hamiltonian_H0}, where the first term is used for moving into the velocity gauge. The resulting Hamiltonian becomes exactly
\begin{equation*}
H_{\rm KH}= \bar{k} + \epsilon\left[\frac{\vert \bar{\bf p}\vert^2}{2}+V\left(\bar{\bf r}+\dfrac{\epsilon^2}{\omega^2} \mathbf{E}(t)\right)\right] + \epsilon^3 \dfrac{\mathbf{E}^2 (t)}{2\omega^2}  .
\end{equation*}
The term of order $\epsilon^3$ in the Hamiltonian $H_{\rm KH}$ can be easily removed by performing an additional transformation. In KH theory, the remaining time-dependence in the potential is removed by an integral over time of the effective potential. For instance, if we expand the KH effective potential up to order $\epsilon^7$ and we average over $t \in [0,T]$, it becomes
\begin{equation*}
V_{\rm KH}=V+\epsilon^4\dfrac{\mathrm{U}_p}{\omega^2 (\xi^2+1)} \left( \dfrac{\partial^2 V}{\partial x^2} + \xi^2 \dfrac{\partial^2 V}{\partial y^2}\right)+O(\epsilon^8).
\end{equation*}
In this framework, the effective potential always only depends on the position variables (and not on the momenta), and there is no contribution at order $\epsilon^6$, contrary to our derivations. The origin of this discrepancy is that performing an averaging using an integral over the fast timescales is only correct to the lowest order (here $\epsilon^4)$, but it fails at higher orders. One needs to perform canonical changes of coordinates to properly average the fast motions. Expressions beyond order $\epsilon^4$ and results obtained using these higher orders are therefore incorrect.   
\par
We hereby take the opportunity to reiterate the advantage of using canonical Lie transforms in the reduction procedure: Since these transformations are invertible and their inverse can be algebraically computed, information on the original system [as described by Hamiltonian~\eqref{eq:main_hamiltonian}] can be fully recovered using the dynamics of the reduced Hamiltonians. The model $G_5$ contains more information than what is provided by KH theory. In particular, the KH theory does not provide $\Phi_5$, and as a consequence, we are not able to reconstruct consistently the trajectory from $H_{\mathrm{KH}}$. 

\section{Analysis of the guiding-center models \label{sec:hierarchy_models}}
In this section, we analyze the different guiding-center models, composed of a time-independent Hamiltonian $H_m$ and a canonical transformation $\Phi_n$. In what follows, we restrict the analysis to a simplified potential for the atoms, namely the soft-Coulomb potential~\cite{Javanainen1988, Becker2012}:
\begin{equation}
\label{eq:Soft_Coulomb_potential}
V({\bf r})=-\frac{1}{\sqrt{\vert {\bf r}\vert ^2+1}}.
\end{equation}

\subsection{Comparison between Hamiltonian~\eqref{eq:main_hamiltonian} and the reduced models}
Figure~\ref{fig:trajectories} shows typical electron trajectories (dark blue curve) of Hamiltonian~\eqref{eq:main_hamiltonian}, the guiding-center trajectories of $\mathrm{G}_2$ (solid light blue curve) and $\mathrm{G}_5$ (solid red curve) and the associated reconstructed trajectories (dashed curves), for $d=2$, $I= 10^{14} \; \mathrm{W}\cdot \mathrm{cm}^{-2}$, $\omega = 0.05 \; \mathrm{a.u.}$ Different ellipticities are considered: $\xi = 0$, $\xi = 0.5$ and $\xi = 1$. The guiding-center trajectories $(\bar{\mathbf{r}}(t), \bar{\mathbf{p}}(t))$ are computed by solving the forward and backward equations of motion of the corresponding guiding-center Hamiltonians~\eqref{eq:Hamiltonians_H2_H5_H7} with initial conditions $\Phi_n (\mathbf{r}(t_0), \mathbf{p}(t_0))$, where $t_0$ is chosen such that $|\mathbf{r} (t_0)| > 50 \; \mathrm{a.u.}$ ($\sim 2 E_0/\omega^2$).
We observe that the light blue and red solid curves guide the oscillating dark blue curves. Therefore, the electron oscillates around a guiding center. For instance, in Figs.~\ref{fig:trajectories}(c)--(d), we see that the trajectory ionizes if the guiding-center motion is unbounded, and it returns to the ionic core if the guiding-center returns to the core. Moreover, we observe a qualitative agreement between the electron trajectory and the reconstructed trajectory using the models $\mathrm{G}_2$ and $\mathrm{G}_5$. In addition, we observe an overlap almost everywhere between the dark blue curve and the red dashed curve, a signature of a very good quantitative agreement between the electron trajectory and the reconstructed trajectory of the model $\mathrm{G}_5$. For the $\mathrm{G_2}$, the overlap is mainly observed far from the ionic core, i.e., for shorter integration times (less than $10T$).  
Below, we provide more thorough analyses to see if and when this agreement between the reduced models and the true trajectories holds. We consider the case $d=1$ for clarity. In what follows, the parameters are $\xi = 0$, $I = 10^{14} \; \mathrm{W}\cdot \mathrm{cm}^{-2}$, and $\omega = 0.05$. 


\begin{figure}
\centering
\includegraphics[width=0.5\textwidth]{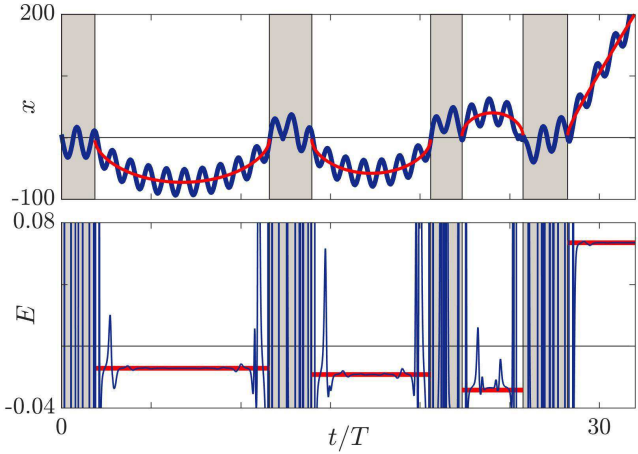}
\caption{Typical electron trajectory of Hamiltonian~\eqref{eq:main_hamiltonian} for $I=10^{14} \; \mathrm{W}\cdot \mathrm{cm}^{-2}$, $d=1$, $\xi=0$ and $\omega = 0.05$, and the guiding-center trajectory of the model $\mathrm{G}_5$, for multiple initial conditions $\Phi (x(t_0), p(t_0))$ such that $|x(t_0)| >50 \; \mathrm{a.u.}$, with forward and backward integration of the equations of motion of $H_5$. The grey areas are where the guiding-center position is $|\bar{x}| < 35 \; \mathrm{a.u.}$ Upper panel: Dark blue and red curves are the electron and the guiding-center trajectory, respectively. Lower panel: The dark blue curve is the guiding-center energy $H_5 (\Phi_5(x(t),p(t)))$. The red curves are the guiding-center energy of $\mathrm{G}_5$, given by $H_5(\Phi_5(x(t_0), p(t_0)))$, respectively. The horizontal black line is $E = 0$. Here $x$ and $E$ are in atomic units.}
\label{fig:Trajectory_1D}
\end{figure}

As a consequence of $d=1$, the electron and guiding-center phase-space coordinates are reduced to $(\mathbf{r} , \mathbf{p}) =  ( x \hat{\mathbf{x}}, p \hat{\mathbf{x}})$ and $(\bar{\mathbf{r}} , \bar{\mathbf{p}}) =  ( \bar{x} \hat{\mathbf{x}}, \bar{p} \hat{\mathbf{x}})$, respectively. Looking at longer trajectories as it is done in Fig.~\ref{fig:Trajectory_1D}, we observe multiple returns of the electron to the ionic core. The upper panel of Fig.~\ref{fig:Trajectory_1D} shows a typical trajectory (dark blue curve) of Hamiltonian~\eqref{eq:main_hamiltonian}, and the guiding-center trajectory for $\mathrm{G}_5$ (red curve) for every interval of time when the electron is far from the ionic core. The guiding-center trajectory is solution of the forward and backward equations of motion of Hamiltonian~\eqref{eq:Hamiltonian_H5}, with initial conditions $\Phi_5 (x(t_0), p(t_0))$, such that $|x(t_0)| > 50 \; \mathrm{a.u.}$ In the lower panel of Fig.~\ref{fig:Trajectory_1D}, the dark blue curve is the guiding-center energy $H_5 (\Phi_5(x(t) , p(t)))$, i.e., at each time, the transformation $\Phi_5$ is performed on the electron phase-space coordinates, and its associated energy $H_5$ is computed. The red curves are the guiding-center energy of Hamiltonian~\eqref{eq:Hamiltonian_H5} for initial conditions $\Phi_5 (x(t_0), p(t_0))$, i.e., the energy of the guiding center of $\mathrm{G}_5$, which is conserved. In Fig.~\ref{fig:Trajectory_1D}, we observe that the guiding center reproduces well the mean trajectory of the electron for several time intervals when the electron is far from the ionic core, in a similar way as it was observed in Fig.~\ref{fig:trajectories}. As a consequence, the guiding-center energy of the electron $H_5 (\Phi_5(x(t) , p(t)))$ is approximately conserved in a piece-wise manner in time. However, we notice that the energy strongly varies during close encounters between the electron and its ionic core. In addition, once the electron has undergone a close encounter, the guiding-center energy of the electron jumps to another energy level.
\par
These observations on the reconstructed trajectories and on the guiding-center energy lead us to consider two different methods for comparing in a more systematic way Hamiltonian~(\ref{eq:main_hamiltonian}) with the $n$-th order guiding-center model $\mathrm{G}_n = (H_m, \Phi_n)$. They consist of:
\begin{enumerate}
\item[(i)]
Computing trajectories of Hamiltonian ${H}_m(\bar{\mathbf{r}},\bar{\mathbf{p}})$ for $t \in [t_0, t_f]$ with initial conditions $\Phi_n(\mathbf{r}(t_0), \mathbf{p}(t_0))$, and then performing the inverse change of coordinates $\Phi_n^{-1}(\bar{\mathbf{r}}(t), \bar{\mathbf{p}}(t))$ to obtain the reconstructed trajectories. 
\item[(ii)]
Computing guiding-center energies ${H}_m (\Phi_n (\mathbf{r}(t), \mathbf{p} (t)))$ with $(\mathbf{r}(t),\mathbf{p}(t))$ the trajectory of Hamiltonian~(\ref{eq:main_hamiltonian}) for $t \in [t_0, t_f]$. 
\end{enumerate}
\par
In what follows, we use these two methods to test the validity and the benefits of the reduced models. Using method (i), the reconstructed trajectories of the model must be close to the true electron trajectory for the models to be relevant. This method is employed in Fig.~\ref{fig:distance_error}. Using method (ii), by definition, the guiding-center energy of the electron ${H}_m (\Phi_n (\mathbf{r}(t), \mathbf{p} (t)))$ must be conserved up to some order for the reduced models to be relevant. This method is employed in Figs.~\ref{fig:rescattering_effect} and~\ref{fig:statistics}. In addition, we use these tests to compare the reduced models and to provide some guidelines on which models should be used for practical purposes. 

\subsubsection{Reconstructed trajectories \label{sec:reconstructed_trajectories}}

\begin{figure}
\centering
\includegraphics[width=0.5\textwidth]{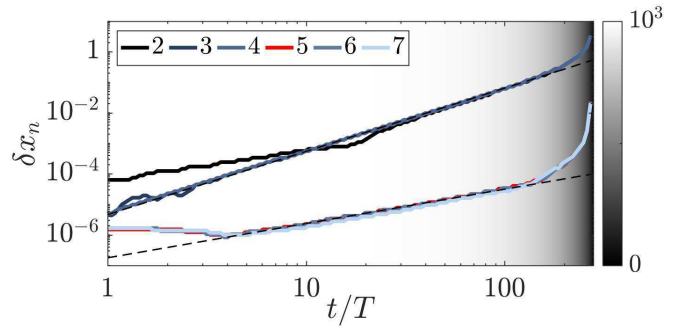}
\caption{Most probable distance error $\delta x_n$ [see Eq.~\eqref{eq:distance_error}] as a function of time $t$ per laser cycle $T$ (in log-log scale) for $I = 10^{14} \; \mathrm{W}\cdot \mathrm{cm}^{-2}$, $d=1$, $\xi=0$ and $\omega = 0.05 \; \mathrm{a.u.}$ The electrons are initialized such that the initial velocity of the guiding center of $\mathrm{G}_2$ is zero, and its initial position is normally distributed, with a mean value $1000 \; \mathrm{a.u.}$ and a standard deviation $5 \; \mathrm{a.u.}$ The initial laser phase is uniformly distributed $\phi \in [0,2\pi]$. The dashed lines are the linear fit curves. The background color is the mean value of the distance of the electron as a function of $t/T$. Here $\delta x_n$ is in atomic units.}
\label{fig:distance_error}
\end{figure}

Figure~\ref{fig:distance_error} shows the most probable distance error $\delta x_n (t)$ between the reconstructed trajectories of the model $\mathrm{G}_n$ and the electron trajectory, such that
\begin{equation}
\label{eq:distance_error}
\delta x_n (t) = | \Pi ( \Phi_n^{-1} ( \bar{x}(t) , \bar{p}(t)) ) - x(t) | ,
\end{equation}
where $x(t)$ is the trajectory of Hamiltonian~\eqref{eq:main_hamiltonian}, $\bar{x}(t)$ is the guiding-center trajectory of Hamiltonian $H_m$ with initial condition $\Phi_n (x(0) , p(0))$, and $\Pi$ is the projection from phase-space onto the position component, i.e., $\Pi (x,p) = x$. The most probable distance error is determined using the maximum of the kernel density estimation~\citep{Sheather1991} of the distance error as a function of $t/T$. Specifically, it is determined in two steps for a fixed $t/T$: First, we compute the kernel density estimation of our data, then, we locate its maximum. The equations of motion for $\bar{x}$ and $\bar{p}$ for the models $\mathrm{G}_2 = (H_2, \Phi_2)$, $\mathrm{G}_3 = (H_2, \Phi_3)$ and $\mathrm{G}_4 = (H_2, \Phi_4)$ are the same, given that the Hamiltonian are the same. The differences between these models come from the change of coordinates, as taken into account in the determination of the initial conditions of the guiding-center trajectory and in the reconstruction of the trajectory from the guiding-center phase-space coordinates. This is also the case when comparing the models $\mathrm{G}_5 = (H_5,\Phi_5)$ and $\mathrm{G}_6 = (H_5, \Phi_6)$.
\par
The distance error between the electron trajectories and the reconstructed trajectories is increasing for increasing time. Moreover, the distance error is increasing faster for $n=\lbrace 2,3,4 \rbrace$ than for $n=\lbrace 5,6,7 \rbrace$. As a consequence, at $t \approx 100 T$, the distance errors $\delta x_n$ for $n=\lbrace 2,3,4 \rbrace$ are two orders of magnitude greater than the ones for $n=\lbrace 5,6,7 \rbrace$. More quantitatively, the most probable distance error $\delta x_n$ scales as
\begin{equation*}
\delta x_n \propto |t|^{\alpha_n} .
\end{equation*}
We observe that $\alpha_n \approx 2.1$ for $n=\lbrace 2,3,4 \rbrace$, and $\alpha_n \approx 1.1$ for $n=\lbrace 5, 6, 7\rbrace$. 
\par
Counter-intuitively, we observe no significant quantitative improvements, neither between the models $\mathrm{G}_2$, $\mathrm{G}_3$ and $\mathrm{G}_4$, nor between the models $\mathrm{G}_5$, $\mathrm{G}_6$ and $\mathrm{G}_7$. Hence, far from the ionic core, the corrective terms in the change of coordinates $\Phi_3$ and $\Phi_4$, are negligible (at least for the chosen parameters), and the models $\mathrm{G}_2$, $\mathrm{G}_3$ and $\mathrm{G}_4$ provide similar results. In the same way, the corrective terms in Hamiltonian~$H_7$, compared with $H_5$, are negligible, as well as the corrective terms in the change of coordinates $\Phi_6$ and $\Phi_7$. 

\subsubsection{Guiding-center energy}


\begin{figure}
\centering
\includegraphics[width=0.5\textwidth]{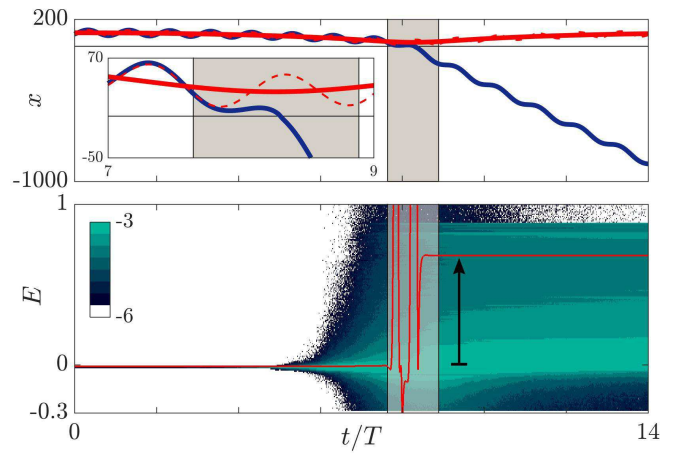}
\caption{Electron trajectory for $I = 10^{14} \; \mathrm{W}\cdot \mathrm{cm}^{-2}$, $d=1$ and $\omega = 0.05 \; \mathrm{a.u.}$ The gray areas are where the guiding-center position is $|\bar{x}| < 35 \; \mathrm{a.u.}$ Upper panel: The dark blue curve is the electron trajectory of Hamiltonian~\eqref{eq:main_hamiltonian} as a function of time $t$ per laser cycle $T$. The solid and dashed red curves are the guiding-center and the approximate trajectory for $\mathrm{G}_5$, respectively, with initial condition $t_0 = 0$. The inset is a zoom of the region around the grey area. Lower panel: Logarithm of the distribution of the guiding-center energy $H_5 (\Phi_5 (x(t), p(t)))$ as a function of $t/T$. The guiding-centers are initialized with a normal distribution with mean value $100 \; \mathrm{a.u.}$ and standard deviation $5 \mathrm{a.u.}$, with zero-velocity and a uniformly distributed initial laser phase $\phi \in [0,2\pi]$. The red curve is the guiding-center energy $H_5 (\Phi_5 (x(t), p(t)))$ of the dark blue curve in the upper panel. Here $x$ and $E$ are in atomic units.}
\label{fig:rescattering_effect}
\end{figure}

We complement the analysis of the trajectories by looking at a specific property of the reduced models, namely the conservation of energy. In Fig.~\ref{fig:rescattering_effect}, an ensemble of trajectories is initiated such that the initial velocity of the guiding center $\mathrm{G}_5$ is zero, and the initial position of the guiding center $\mathrm{G}_5$ is normally distributed with a mean value $100 \; \mathrm{a.u.}$ and a standard deviation $5 \; \mathrm{a.u.}$ The initial laser phase is uniformly distributed $\phi \in [0,2\pi]$. The distribution in the lower panel of Fig.~\ref{fig:rescattering_effect} represents the distribution of the guiding-center energy $H_5 (\Phi_5(x(t) , p(t)))$ as a function of $t/T$. In the upper panel, the dark blue curve is a typical electron trajectory in the ensemble. The red curve is the guiding-center energy $H_5 (\Phi_5 (x(t),p(t)))$ of the electron trajectory corresponding to the dark blue curve in the upper panel.  
\par
For $t < 7 T$, we observe in the lower panel that the distribution is peaked around the initial guiding-center energy of the electron. During this time, the electrons are far from the ionic core. At $t \sim 8 T$, the electrons get close to the ionic core, and the guiding-center energy distribution starts to spread out.  When the electrons are close to the ionic core, their dynamics is highly nonlinear due to the competition between the strong laser and Coulomb fields, and the energy curve of a single trajectory in the lower panel starts varying significantly. In the meantime, the inset in the upper panel shows that the approximate trajectory of the reduced model no longer reproduces the electron trajectory. For $t > 9 T$, the electron is far from the ionic core, and the red curve in the lower panel stops varying. It means that the model $\mathrm{G}_5$ is again relevant, but for a different energy level than the initial energy. The arrow indicates the jump of the guiding-center energy after the close encounter with the ionic core. Close encounters with the ionic core are short time processes, and therefore cannot be averaged in time. It is expected that the fast-time average we perform fails to describe the various energy exchanges happening on these short time scales.


\begin{figure}
\centering
\includegraphics[width=0.5\textwidth]{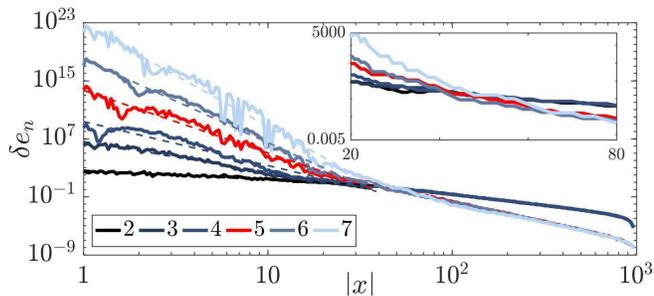}
\caption{Most probable energy error $\delta e_n$ [see Eq.~\eqref{eq:energy_error}] as a function of the distance between the electron and the ionic core $|x|$ (in log-log scale), for $I = 10^{14} \; \mathrm{W}\cdot \mathrm{cm}^{-2}$, $d=1$ and $\omega = 0.05 \; \mathrm{a.u.}$ The initial conditions are the same as in Fig.~\ref{fig:distance_error}. The dashed lines are the linear approximation for $|x| \in [1, 35] \; \mathrm{a.u.}$ The inset shows a zoom of the curves. Here $|x|$ is in atomic units.}
\label{fig:statistics}
\end{figure}

Figure~\ref{fig:statistics} shows the most probable relative energy error $\delta e_n (t)$ for the models $\mathrm{G}_n$ as a function of the distance between the electron and the ionic core $|x(t)|$, such that
\begin{equation} 
\label{eq:energy_error}
\delta e_n (t) = \left| \dfrac{H_m (\Phi_n (x(t),p(t))) - H_m (\Phi_n (x(0),p(0)))}{H_m (\Phi_n (x(0),p(0))) } \right| ,
\end{equation} 
where $(x(t), p(t))$ are the electron phase-space coordinates at time $t$. The most probable energy error is the maximum of the kernel density estimation~\citep{Sheather1991} of the energy error. It is determined using the same technique as for computing the most probable distance error (see Fig.~\ref{fig:distance_error}). The initial conditions are the same as in Fig.~\ref{fig:distance_error}, and the integration is stopped when the electron reaches $x = 1 \; \mathrm{a.u.}$
\par
As expected, we observe that the energy error $\delta e_n$ increases when $|x|$ decreases, i.e., as the electron approaches the ionic core. Far away from the ionic core, we observe that the most probable energy error $\delta e_n$ scales as
\begin{equation*}
\delta e_n \propto |x|^{- \beta_n} ,
\end{equation*}
with $\beta_n \approx 3.0$ for $n=\lbrace 2, 3, 4 \rbrace$, and $\beta_n \approx 6.5$ for $n=\lbrace 5, 6, 7\rbrace$. As in Sec.~\ref{sec:reconstructed_trajectories}, for we observe no significant quantitative improvements, neither between the models $\mathrm{G}_2$, $\mathrm{G}_3$ and $\mathrm{G}_4$, nor between the models $\mathrm{G}_5$, $\mathrm{G}_6$ and $\mathrm{G}_7$.
\par
A cross-over is observed between all the models when the electron reaches $\sim 35 \; \mathrm{a.u.}$ 
In particular, from Fig.~\ref{fig:statistics}, we observe that $\mathrm{G}_2 = (H_2, \Phi_2)$ gives the smallest energy errors among the reduced models close to the ionic core. The main reason is that $H_2$ and $\Phi_2$ do no contain derivatives of the potential. As such, $\mathrm{G}_2$ constitutes the most robust model among the hierarchy. Far from the ionic core ($>35\; \mathrm{a.u.}$), the higher order models provide a better quantitative agreement with the electron trajectories as shown in Figs.~\ref{fig:distance_error} and~\ref{fig:statistics}. The efficiency of the higher-order models appears far from the ionic core, around $35 \; \mathrm{a.u.}$

\subsubsection{Discussion \label{sec:Discussion}}

\begin{figure}
\centering
\includegraphics[width=0.5\textwidth]{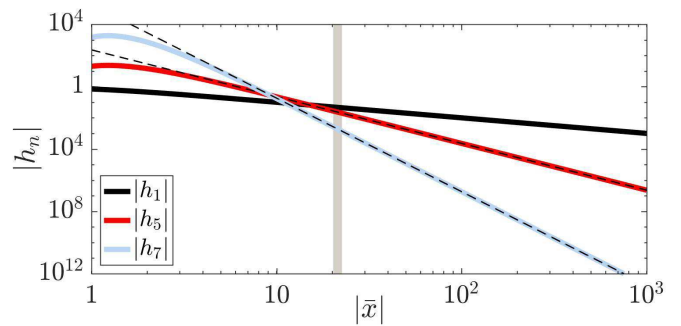}
\caption{Representation of $h_n$ as a function of the distance between the guiding center and the ionic core $|\bar{x}|$ (in log-log scale) for $I = 10^{14} \; \mathrm{W}\cdot \mathrm{cm}^{-2}$, $d=1$ and $\omega = 0.05$. The term $h_n$ corresponds to the term of order $O(\epsilon^n)$ in the Hamiltonian~\eqref{eq:total_averaged_Hamiltonian}, which are: $h_1 = V(\bar{x})$, $h_5 = (\mathrm{U}_p/\omega^2) V^{\prime\prime}(\bar{x})$, and $h_7 = (\mathrm{U}_p/\omega^4) V^{\prime\prime}(\bar{x})^2$. The dashed lines are the asymptotic behaviors of these terms for soft-Coulomb potential~\eqref{eq:Soft_Coulomb_potential}, which are for $|x| \gg 1$: $|h_1| \propto |\bar{x}|^{-1}$, $|h_5| \propto |\bar{x}|^{-3}$ and $|h_7| \propto |\bar{x}|^{-6}$. The vertical line is $x_c = E_0/\omega^2$. Here $h_n$ and $|\bar{x}|$ are in atomic units.}
\label{fig:h_n_potentials}
\end{figure}

The relative failure of the higher order models close to the ionic core can be understood by looking at the magnitude of the corrective terms in the Hamiltonians $H_m$ and in the changes of coordinates $\Phi_n$. Figure~\ref{fig:h_n_potentials} shows $h_1 = V(\bar{x})$, $h_5= (\mathrm{U}_p/\omega^2) V^{\prime\prime}(\bar{x})$ and $h_7= (\mathrm{U}_p/\omega^4) V^{\prime\prime}(\bar{x})^2$ as a function of the distance between the guiding center and the ionic core. The term $h_n$ corresponds to the time-independent term in Hamiltonian~\eqref{eq:total_averaged_Hamiltonian} of order $\epsilon^{n}$. We observe an overlap between these terms, where $h_1 \sim h_5$ for $|\bar{x}| \approx x_c$. Far from the ionic core, we approximate the soft-Coulomb potential by a hard-Coulomb potential $V(\bar{x}) \approx - 1 / |\bar{x}|$ and we compute $h_n$ explicitly. We deduce that $h_1 \sim h_5$ for $\bar{x} \approx x_c$ with
\begin{equation*}
x_c \sim E_0 / \omega^2 ,
\end{equation*}
which is approximately equal to $21 \; \mathrm{a.u.}$ for $I = 10^{14} \; \mathrm{W} \cdot \mathrm{cm}^{-2}$ and $\omega = 0.05 \; \mathrm{a.u.}$ Therefore, for $|\bar{x}| \gg x_c$, the terms are ordered such that $h_1 > h_5 > h_7$. In this case, the series in the perturbative expansion of Hamiltonian~\eqref{eq:total_averaged_Hamiltonian} and the change of coordinates~\eqref{eq:total_coordinate_transformation} are likely converging, and the models $\mathrm{G}_n$ for $n>2$ are relevant for the guiding-center dynamics. For $|\bar{x}| \ll x_c$, the terms are ordered such that $h_1 < h_5 < h_7$. In this case, the series in the perturbative expansion of Hamiltonian~\eqref{eq:total_averaged_Hamiltonian} and the change of coordinates~\eqref{eq:total_coordinate_transformation} are likely diverging, and the models $\mathrm{G}_n$ for $n>2$ are no longer relevant for the dynamics of Hamiltonian~\eqref{eq:main_hamiltonian}.

\subsection{Guiding-center phase-space dynamics}
\subsubsection{For $H_2$}
The first (non-trivial) element of the hierarchy is $\mathrm{G}_2 = (H_2,\Phi_2)$. This model was identified above as the most robust one in the hierarchy for the analysis of the guiding-center dynamics, since its guiding-center energy error is lower than for any other models close to the ionic core. We notice that this is the only reduced Hamiltonian which does not depend on the parameters of the laser field. The dependence on the laser field is in the change of variables $\Phi_2$. Moreover, if the potential is rotationally invariant, as is the case for atoms, the resulting Hamiltonian is integrable since the angular momentum is conserved in addition to the Hamiltonian. 
\par
The change of variables is exactly given by
\begin{eqnarray*}
\bar{\mathbf{r}} &=&  \mathbf{r} - \epsilon^2 \mathbf{E}(t)/ \omega^2 , \\
\bar{\mathbf{p}} &=& \mathbf{p} - \epsilon \mathbf{A}(t). 
\end{eqnarray*}
What is particularly convenient with this guiding-center model is that the potential is taken into account in the Hamiltonian and the electric field in the change of variables. 
\par
Ionization occurs if and only if the energy of the guiding center $E=H_2 (\bar{\mathbf{r}}(t),\bar{\mathbf{p}}(t))$ is positive. Otherwise the motion of the electron is bounded since the guiding center moves on a quasi-periodic orbit. The laser parameters have no influence on the motion of the guiding centers (and this holds up to the fourth-order model). They only influence how the electron swirls around the quasi-periodic orbit.
\par
We consider the case when the guiding-center is far from the ionic core and as a consequence we can approximate $V(\bar{\mathbf{r}}) \approx -1/|\bar{\mathbf{r}}|$. Thus, the typical guiding-center trajectory is on a Kepler orbit for $E<0$, where $E$ is the energy of the orbit, as it is the case for the guiding-center trajectories in Figs.~\ref{fig:trajectories}(a--c) and~(f). One of the particularities of these orbits is that the radial momentum $\bar{\mathbf{p}}\cdot \bar{\mathbf{r}}/|\bar{\mathbf{r}}|$ vanishes twice in a revolution cycle: Once when the guiding-center trajectory is at the perihelion $r$ (minimum distance from the ionic core) and $\mathrm{d}(\bar{\mathbf{p}}\cdot \bar{\mathbf{r}}/|\bar{\mathbf{r}}|)/\mathrm{d}t > 0$, and once when the guiding-center trajectory is at the aphelion $R$ (maximum distance from the ionic core) and $\mathrm{d}(\bar{\mathbf{p}}\cdot \bar{\mathbf{r}}/|\bar{\mathbf{r}}|)/\mathrm{d}t < 0$. The aphelion and the perihelion are such that $R + r = 1/|E|$, imposing that for a given energy, the larger the aphelion, the smaller the perihelion, i.e., the closer the electron gets to the ionic core. 

\subsubsection{For $H_5$ and $H_7$}

\begin{figure*}
\centering
\includegraphics[width=\textwidth]{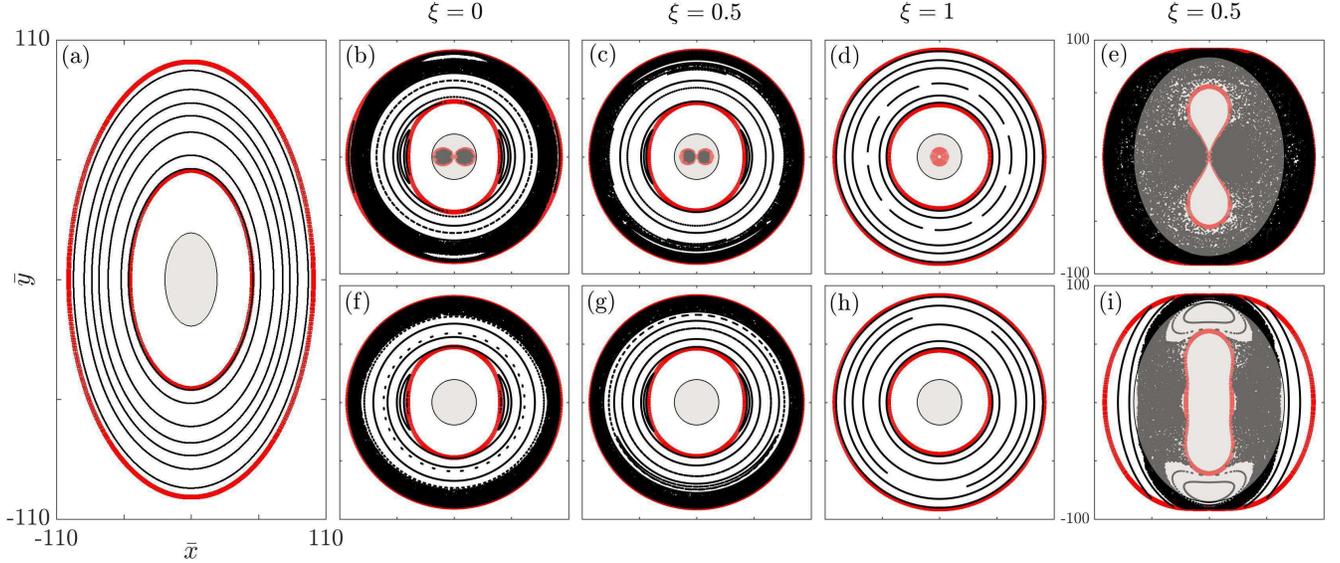}
\caption{Poincar\'{e} sections $\bar{\mathbf{r}}\cdot \bar{\mathbf{p}} = 0$ and $\mathrm{d} (\bar{\mathbf{p}} \cdot \bar{\mathbf{r}}/|\bar{\mathbf{r}}|)/\mathrm{d}t < 0$ in the polarization plane $(\bar{x},\bar{y})$, for potential~\eqref{eq:Soft_Coulomb_potential}, $I = 1\times 10^{14} \; \mathrm{W.cm}^{-2}$ and $E = -0.01 \; \mathrm{a.u.}$ The grey areas are $|\mathbf{r}| < E_0 / \omega^2$. The Hamiltonians are: (a) $H_2$, (b--e) $H_5$, and (f--i) $H_7$. The ellipiticities are: (b,f) $\xi = 0$, (c,e,g,i) $\xi = 0.5$, (d,h) $\xi = 1$. The frequencies are $\omega = 0.05$ except for (d,h) $\omega = 0.025$ (larger quiver radius). All axes are the same as for (a) unless stated otherwise. All quantities are in atomic units.}
\label{fig:Poincare_sections}
\end{figure*}

When going to higher-order models, the Hamiltonian $H_2$ gets perturbed by $h_5$ (and $h_7$). As a consequence, since the perturbation mainly affects the trajectories that pass close to the ionic core, i.e., $r < x_c \sim E_0/\omega^2$, we shall see that the most perturbed trajectories in the higher-order models are the ones with a large aphelion $1/|E| - E_0/\omega^2 < R < 1/|E|$. 
\par
The fact that the energy of the guiding centers is conserved is a property which is preserved by construction of the reduction procedure: $E = H_5 (\bar{\mathbf{r}}(t),\bar{\mathbf{p}}(t))$ or $E=H_7 (\bar{\mathbf{r}}(t),\bar{\mathbf{p}}(t))$ is conserved in time. Consequently, for $d=2$, the dimension of phase space is reduced from $5$ to $3+1$. Here $3+1$ means that phase space is foliated by constant energy surfaces of dimension $3$. The advantage is that one can visualize the dynamics using Poincar\'e sections. Figure~\ref{fig:Poincare_sections} shows the Poincar\'{e} sections $\bar{\mathbf{r}}\cdot \bar{\mathbf{p}} = 0$ and $\mathrm{d} (\bar{\mathbf{p}} \cdot \bar{\mathbf{r}}/|\bar{\mathbf{r}}|)/\mathrm{d}t < 0$ for $E=-0.01 \; \mathrm{a.u.}$ and the soft-Coulomb potential~\eqref{eq:Soft_Coulomb_potential}, where $\bar{\mathbf{p}}\cdot\bar{\mathbf{r}}/|\bar{\mathbf{r}}| = \mathrm{d}|\bar{\mathbf{r}}|/\mathrm{d}t$ is the radial momentum of the electron. This Poincar\'e section corresponds to the position of the guiding-center when it turns back towards the ionic core. The region close to the ionic core, i.e., around $|\bar{\mathbf{r}}| \sim E_0/\omega^2$ (grey areas), is not relevant since the reduction procedure is only valid far away from the ionic core. 
\par
For linear polarization ($\xi = 0$), in Figs.~\ref{fig:Poincare_sections}(b) and~(f), we observe two distinct dynamical behaviors of the guiding-center trajectories. These figures display a chaotic layer far from the ionic core around $1/|E| - E_0/\omega^2 < |\bar{\mathbf{r}}| < 1/|E|$. This ring corresponds to trajectories with a small perihelion and a large aphelion, that come close to the ionic core $|\bar{\mathbf{r}}| < E_0/\omega^2$. These trajectories are the most affected by the perturbation $h_5$ in the Hamiltonian $H_5$ according to our discussion in Sec.~\ref{sec:Discussion}, and typically correspond to the trajectories for which the electron comes back to the ionic core, as it is depicted in Figs.~\ref{fig:trajectories}(a),(c) and~(f). The width of this ring is of order $E_0/\omega^2$. Secondly, we observe a regular region for $1/2|E| < |\bar{\mathbf{r}}| < 1/|E| - E_0/\omega^2$. This region corresponds to trajectories with a perihelion greater than $E_0/\omega^2$, of the same order of their aphelion. These are the trajectories least affected by the perturbations $h_5$ and $h_7$. As a consequence, we observe orbits that are mostly preserved from the unperturbed Hamiltonian $H_2$, and typically correspond to the trajectories for which the electron stays far from the ionic core, as it is depicted in Fig.~\ref{fig:trajectories}(b). Also, we observe two elliptic islands for $(\bar{x} , \bar{y}) \sim (\pm 1/2|E|, 0)$. These islands correspond to nearly circular guiding-center orbits, which become stable with the coupling with the electric field encapsulated in the effective potential of $H_5$ or $H_7$. Finally, the Poincar\'e sections for $H_5$ and $H_7$ look similar. This is another indication that $h_7$ does not bring enough perturbation compared to $h_5$ for these parameters and far from the ionic core. 
\par
For elliptical polarization ($\xi = 0.5$), in Figs.~\ref{fig:Poincare_sections}(c) and~(g), the observations are similar to the linear case, which reinforces the generality of the discussion above. However, for circular polarization ($\xi = 1$), in Figs.~\ref{fig:Poincare_sections}(d) and~(h), we no longer observe chaotic behavior in the guiding-center dynamics. Indeed, Hamiltonians $H_5$ and $H_7$ are rotationally invariant, and as a consequence, the guiding-center angular momentum is conserved. Therefore, the dimension of phase space is reduced from $3+1$ to $2+2$ and the system is integrable. We observe that the elliptic islands we observed for $\xi = 0$ and $0.5$ around $(\bar{x} , \bar{y}) \sim (\pm 1/2|E|, 0)$ are no longer present in the circular polarization case, as a consequence of the rotational invariance. 
\par
In the previous sections we have seen that for the parameters $I=10^{14} \; \mathrm{W}\cdot \mathrm{cm}^{-2}$ and $\omega = 0.05 \; \mathrm{a.u.}$, the models $H_5$ and $H_7$ are almost equivalent since $h_5$ is several order of magnitude higher than $h_7$ in the region where the higher order models are relevant, $|\bar{\mathbf{r}}| > E_0/\omega^2$ (see Fig.~\ref{fig:h_n_potentials}). This is verified by comparing the Poincar\'e sections of Figs.~\ref{fig:Poincare_sections}(b--d) and Figs.~\ref{fig:Poincare_sections}(f--h). However, in Figs.~\ref{fig:Poincare_sections}~(e)--(i), we observe that when $\omega = 0.025 \; {\rm a.u.}$ so that the quiver radius is of the same order as the distance between the guiding-center and the ionic core ($E_0/\omega^2 \sim 85 \; \mathrm{a.u.}$), the dynamics between Hamiltonians $H_5$ and $H_7$ differs significantly. The reduced models are not relevant in these regions. Hence, the reduced models $\mathrm{G}_5$ and $\mathrm{G}_7$ are significantly different when the characteristic distance between the guiding center and the ionic core is the same as the quiver radius (at least for this range of parameters). Similar observations would have been made for $\omega = 0.05 \; \mathrm{a.u.}$ by lowering the guiding-center energy $E$.

\section*{Conclusions}

We have derived a hierarchy of reduced models $\mathrm{G}_n$ for the guiding-centers dynamics of the electron interacting with the combined strong laser and Coulomb fields. The reduced models $\mathrm{G}_n$ are composed of an averaged Hamiltonian $H_m$ governing the guiding-center dynamics [Eqs.~\eqref{eq:Hamiltonians_H2_H5_H7}] and a transformation $\Phi_n$ which maps the electron phase-space coordinates onto the guiding-center phase-space coordinates [Eqs.~\eqref{eq:transformation_Phi_n}]. As a rule of thumb, these models are relevant when the electron is relatively far away from the ionic core (typically when its distance from the core exceeds one quiver radius), which happens in a piece-wise manner in time. The models do not describe the short events when the electron recollides with the ionic core. 
\par
We have singled out two models $\mathrm{G}_2$ and $\mathrm{G}_5$: The first model provides the leading behavior of the trajectories and is the most tractable one due to its simplicity. In order to improve the quantitative agreement, a higher-order model such as $\mathrm{G}_5$ has to be used.   
\par
All these models allow the distinction between direct ionizations and rescattering with the ionic core. This is  very useful when the photoelectron momentum distributions are analyzed for imaging the target. In particular, we were able to define an energy of the electron far away from the core for this time-dependent system. The rescattering events can be seen as jumps in energy as a result of the transfer of energy from the ionic core to the electron. 

\section*{Acknowledgments}
The project leading to this research has received funding from the European Union's Horizon 2020 research and innovation program under the Marie
Sklodowska-Curie grant agreement No. 734557. T.U. and S.A.B. acknowledge funding from the NSF (Grant No. PHY1304741).


%

\end{document}